\documentclass[aps,twocolumn,tightenlines,nofootinbib,superscriptaddress,floatfix]{revtex4-2}
\usepackage[utf8]{inputenc}
\usepackage{graphicx}
\usepackage{subcaption}
\usepackage{mwe}
\usepackage{mathrsfs}
\usepackage{amsfonts}
\usepackage{setspace}
\usepackage{cellspace}
\usepackage{amsmath,amssymb,bm}
\usepackage[colorlinks=true,linkcolor=blue]{hyperref}
\usepackage{xcolor}
\usepackage{epsfig}
\usepackage{slashed}
\usepackage{caption}
\usepackage{hhline,multirow,tabularx}  %
\usepackage{dcolumn}    %
\usepackage{url}        %
\usepackage{braket}     %
\usepackage{makecell}
\usepackage{bbm}
\usepackage{subcaption} 
\usepackage[normalem]{ulem}

\newcommand{\dd}{\text{d}}

\DeclareMathAlphabet{\mathdutchcal}{U}{dutchcal}{m}{n}
\SetMathAlphabet{\mathdutchcal}{bold}{U}{dutchcal}{b}{n}
\DeclareMathAlphabet{\mathdutchbcal}{U}{dutchcal}{b}{n}

\begin{document}

\title{Benchmarking the Nearside Energy-Energy Correlators with Mellin Transform}

\author{Yuxun Guo}
\email{yuxunguo@lbl.gov}
\affiliation{Nuclear Science Division, Lawrence Berkeley National
Laboratory, Berkeley, CA 94720, USA}
\affiliation{Physics Department, Brookhaven National Laboratory, Upton, NY 11973, USA}
\affiliation{RIKEN BNL Research Center, Brookhaven National Laboratory, Upton, NY 11973, USA}

\author{Feng Yuan}
\email{fyuan@lbl.gov}
\affiliation{Nuclear Science Division, Lawrence Berkeley National
Laboratory, Berkeley, CA 94720, USA}

\begin{abstract}
We investigate nearside energy-energy correlators (EECs) at small angles, explicitly incorporating the QCD scaling behavior in both the perturbative and post-confinement regimes through a Mellin-transform framework.
As an illustration, we show that a single parameter, $\Lambda$, characterizing the transition scale between the two regimes, provides an excellent description of nearside EECs in $e^+e^-$ annihilation, including the recent ALEPH analysis as well as earlier measurements across different energies, %
with next-to-next-to-leading-order accuracy and next-to-next-to-leading-logarithmic resummation.
\end{abstract}
 
\maketitle

\section{Introduction}

Energy-Energy Correlators (EECs) of particle productions in $e^+e^-$ annihilation and in jets in hadronic collisions have played an important role in studying the strong interaction physics at the collider frontier~\cite{Basham:1978bw,Basham:1977iq,Basham:1978zq,PLUTO:1985yzc,PLUTO:1979vfu,CELLO:1982rca,JADE:1984taa,Fernandez:1984db,Wood:1987uf,TASSO:1987mcs,AMY:1988yrv,TOPAZ:1989yod,ALEPH:1990vew,L3:1991qlf,L3:1992btq,DELPHI:1990sof,OPAL:1990reb,OPAL:1991uui,SLD:1994idb,Collins:1981uk,Ali:1982ub,Clay:1995sd,deFlorian:2004mp,DelDuca:2016csb,Tulipant:2017ybb,Kardos:2018kqj,Moult:2018jzp,Dixon:2018qgp,Dixon:2019uzg,Ebert:2020sfi,Schindler:2023cww,Lee:2006nr,Berger:2003iw,Hofman:2008ar,Chen:2020vvp,Lee:2022ige,Craft:2022kdo,Komiske:2022enw,Andres:2022ovj,Andres:2023xwr,Andres:2023ymw,Yang:2023dwc,Andres:2024ksi,Barata:2023bhh,Barata:2023zqg,Bossi:2024qho,Lee:2023tkr,Lee:2023xzv,Lee:2024esz,Chen:2024nyc,Holguin:2023bjf,Holguin:2024tkz,Xiao:2024rol,Xing:2024yrb,Andres:2024hdd,Liu:2024lxy,Alipour-fard:2024szj,Kang:2024dja,Cuerpo:2025zde,Barata:2024wsu,Csaki:2024zig,Fu:2024pic,Apolinario:2025vtx,Barata:2025fzd,Chen:2025rjc,Moult:2025nhu,CMS:2024mlf,ALICE:2024dfl,Tamis:2023guc,CMS:2025ydi,ALICE:2025igw,ALICEpA,Barata:2024ukm,Lee:2025okn,Chang:2025kgq,Guo:2025zwb,Kang:2025zto,Guo:2025qnz,Herrmann:2025fqy,Zhao:2025ogc,Cao:2025icu,Electron-PositronAlliance:2025fhk,Jaarsma:2025tck,Song:2025bdj,Csaki:2025abk,Holguin:2026vld,Kang:2026hig,Ren:2026cvu,Chen:2026hmd,Barata:2026pgh,Kang:2026pro,Yang:2026bsk}. 
Renewed interests in recent years have led to a suite of new measurements across different colliders, including the Relativistic Heavy Ion Collider, the Large Hadron Collider, as well as proposals for measurements at the planned  Electron-Ion Collider (EIC). They not only provide an opportunity to explore quantum chromodynamics (QCD) and shed light on the hadronization effects in high-energy processes, but also lead to unique opportunities to explore strong interaction physics in different aspects, for example, the applications in heavy ion collisions %
and hadron spin physics; %
see, a recent review~\cite{Moult:2025nhu}.

Two important features have been unveiled from recent theoretical developments. First, the connection to formal field theory not only demonstrates the interplay of EECs physics with formal field theory~\cite{Hofman:2008ar, Kravchuk:2018htv, Kologlu:2019mfz} and perturbative QCD~\cite{Moult:2018jzp, Dixon:2019uzg, Chen:2023zzh} but also provides deep insight into distinct behaviors across different kinematic regions of EECs. Especially, the classical and quantum scalings can be derived directly from this correspondence~\cite{Chang:2025kgq}. %

Meanwhile, a connection between the nearside EECs and the di-hadron fragmentation functions~\cite{Konishi:1978yx,Konishi:1979cb,Vendramin:1981te,deFlorian:2003cg,Majumder:2004wh,Majumder:2004br,Jaffe:1997hf,Bianconi:1999cd,Bacchetta:2002ux,Bacchetta:2012ty,Zhou:2011ba,Cocuzza:2023vqs,Pitonyak:2023gjx,Rogers:2024nhb} has been made in~\cite{Lee:2025okn,Chang:2025kgq,Guo:2025zwb,Guo:2025qnz,Kang:2025zto} which provides an unique method to explore the particle hadronization at high energy colliders. In particular, %
one can introduce the EEC jet functions, which can be constructed from di-hadron fragmentation functions, summing over all hadrons weighted with their energy fractions. This connection helps extend the factorization formula derived in~\cite{Dixon:2019uzg} to the non-perturbative region where di-hadron fragmentation function contributions dominate. 

In previous studies, EECs were computed either through direct numerical or iterative convolutions~\cite{Lee:2025okn, Herrmann:2025fqy, Kang:2025zto} or approximated with analytic insights~\cite{Guo:2025zwb, Guo:2025qnz}.
Here, we present a Mellin-transform-based framework that incorporates perturbative corrections and scale evolutions systematically and rigorously. The Mellin transform is a well-established and powerful technique widely applied in QCD phenomenology, where parton interactions naturally appear as Mellin convolutions~\cite{Vogt:2004ns, deFlorian:2007aj, deFlorian:2014yva, Borsa:2022vvp, Borsa:2024mss, Mueller:2005ed, Kumericki:2007sa, Kumericki:2009uq, Guo:2022upw, Guo:2023ahv, Guo:2025muf}.

To illustrate this, we focus on the EECs in $e^+e^-$ annihilation, which can be fully generalized to EECs in jets as well. The EECs in $e^+e^-$ annihilation are defined as
\begin{equation}
\frac{\dd\Sigma_2^{e^+e^-}}{\dd\cos\chi} = \frac{1}{\sigma_{\text{tot}}} \sum_{i,j} \int \dd\sigma^{ij} \, \frac{E_i E_j}{Q^2} \, \delta\left(\cos\chi - \hat{n}_i \cdot \hat{n}_j\right),\label{eq:eecdefg}
\end{equation}
where $\dd\sigma^{ij}$ denotes the semi-inclusive differential cross section for producing two hadrons $h_i$ and $h_j$ with energies $E_i$ and $E_j$, moving along the unit directions $\hat{n}_i$ and $\hat{n}_j$, respectively. The total energy is $Q=\sum_i E_i$, and $\chi$ denotes the angle between the two directions. For convenience, we define the variable $z \equiv (1 - \cos\chi)/2$. In the nearside region with small $z\ll 1$, the EEC is dominated by collinear gluon radiations, and a QCD factorization has been established~\cite{Dixon:2019uzg}. %

Within the Mellin-transform framework, we incorporate higher-order corrections up to next-to-next-to-leading order (NNLO) and next-to-next-to-leading logarithmic (NNLL) resummation. Compared to previous studies at the same theoretical accuracy but using iterative solutions~\cite{Herrmann:2025fqy,Jaarsma:2025tck}, this approach allows us to implement two key boundary conditions motivated by prior studies. First, we adopt an initial-scale parameterization that ensures consistent perturbative scaling at large transverse momentum $q_T=\sqrt{z}Q$~\cite{Dixon:2019uzg}. Second, we incorporate constraints from di-hadron fragmentation functions into the initial input. Both conditions are naturally implemented in $b_T$ space~\cite{Guo:2025zwb, Guo:2025qnz}, and the distinct scaling behaviors in different kinematic regions are guaranteed with the Mellin-space derivations; see below.

The results, and especially the methodology developed in this work, establish a benchmark for nearside EEC studies at colliders, which serves as a crucial component for precision studies to extract the running coupling constant, determine hot QCD matter properties, and investigate nucleon tomography from future EEC measurements at the EIC. More importantly, the close connection between our method in Mellin transform space and the formal theory analysis of light-ray operator product expansion (OPE) shall stimulate further developments to shed light on the underlying physics of the nearside EECs in various hadronic collisions.

\section{EEC Jet Functions in Mellin Space}

In the collinear region of EECs in $e^+e^-$ annihilation, we can write down the factorization in either $q_T$ or $b_T$ space. For example, in $b_T$ space, we have~\cite{Guo:2025qnz}, 
\begin{equation}
    \left.2\frac{\text{d}\Sigma_2^{e^+e^-}}{\pi Q^2\text{d} z}\right|_{z\ll 1}={\sigma_0\over\sigma_{\textrm{tot}}}
    \int \frac{b_T\text{d}b_T}{(2\pi)}J_0\left({\sqrt{z}Qb_T}\right)\widetilde{\Sigma}^{e^+e^-}_2(Q,b_T) \ ,
\end{equation}
and the factorization of $\widetilde{\Sigma}_2$ is similar to that in~\cite{Dixon:2019uzg},
\begin{equation}
    \widetilde{\Sigma}^{e^+e^-}_2(Q,b_T)=\int {\text{d}x}\, {x^2}H^{e^+e^-}_i\left(x,{Q\over \mu}\right)\widetilde{\Gamma}_i\left(\mu,{b_T\over x}\right) \ .\label{eq:factorizationbtee}
\end{equation}
The $H^{e^+e^-}_i(x,Q/\mu)$ denotes $H_i^{(e^+e^-\to hX)}(x,Q/\mu)$, the hard coefficients of parton $i$ in the single inclusive hadron production in $e^+e^-$ annihilation, which has been computed in the literature~\cite{Altarelli:1979kv, deFlorian:2003cg, Rijken:1996ns, Mitov:2006wy, He:2025hin}. The $\widetilde{\Gamma}_i(\mu,b_T)$ refers to the unintegrated EEC jet function of parton $i$ in $b_T$ space, related to the $q_T$-space unintegrated EEC jet function $\Gamma_i(\mu,q_T)$ by Fourier transform. Its scale $\mu$ dependence satisfies the time-like collinear DGLAP evolution,  
\begin{eqnarray}
    \frac{\partial \widetilde{{\Gamma}}_i(\mu,b_T)}{\partial\ln\mu^2}&=&\int{\dd x}\,  x^2  \hat P_{T,ji}(x) \widetilde{{\Gamma}}_j\left(\mu,\frac{b_T}{x}\right) \ ,\label{eq:evolutionbt}
\end{eqnarray}
with $\hat P_{T,ji}(x)$ the time-like DGLAP splitting kernel.

The Mellin convolution form of the factorization and evolution equations (\ref{eq:factorizationbtee}) and (\ref{eq:evolutionbt}) indicates that they become multiplicative in Mellin space, as dictated by the light-ray OPE analysis as well~\cite{Chen:2024nyc, Chang:2025kgq}. Thus, we introduce the Mellin moments of the $b_T$-space EEC jet functions with respect to $b_T$:
\begin{eqnarray}
\widetilde{\Gamma}_i(\mu,N)&\equiv&\int_0^\infty\text{d} b_T\ b_T^{N-1} \widetilde{\Gamma}_i(\mu,b_T)\ , 
\end{eqnarray}
and those of the splitting kernel with respect to $x$:
\begin{eqnarray}
\gamma^{(N)}_{T,ji}&\equiv&-\int_0^1\text{d}x \,x^{N-1} \hat P_{T,ji}(x)\ ,
\end{eqnarray}
also known as the time-like anomalous dimensions. Then the evolution equation becomes
\begin{eqnarray}
    \frac{\partial {\widetilde{\Gamma}}_i(\mu,N)}{\partial\ln\mu^2}&=&\int{\dd x}\,  x^{2+N} \hat P_{ji}(x) \widetilde{\Gamma}_j(\mu,N) ,\nonumber \\
    &=&- \gamma^{(3+N)}_{ji}\widetilde{\Gamma}_j(\mu,N)\ .
\end{eqnarray}
For simplicity, henceforward, we suppress the $T$ subscript of the time-like splitting kernels and anomalous dimensions, which are distinct from the space-like ones starting at next-to-leading order,  and implicitly assume them to be time-like unless otherwise specified.

The multiplicative form of the evolution equation admits a simple exponential solution,
\begin{equation}
    \widetilde\Gamma_i(\mu,N)=\widetilde\Gamma_j(\mu_0,N)\, \mathcal{E}_{ji}(\mu,\mu_0;N+3)\ ,
\end{equation}
with evolution operator $\mathcal{E}_{ji}(\mu,\mu_0;N)$ defined as,
\begin{equation}
    \mathcal{E}_{ji}(\mu,\mu_0;N) \equiv \text{P}\exp\left[-\int_{\mu_0^2}^{\mu^2}\frac{\text{d}\mu^{\prime 2}}{\mu^{\prime 2}} \gamma^{(N)}(\mu')\right]_{ji}\ , \label{eq:evolmatrix}
\end{equation}
where $\text{P}$ implies a path order of the integral variable $\mu'$.

Then the $b_T$-space unintegrated EEC jet functions evolved to an arbitrary scale $\mu$ can be obtained via the inverse Mellin transform, expressed as a contour integral:
\begin{equation}
    \label{eq:mellinevobT}
    \widetilde{\Gamma}_i(\mu,b_T)=\int\limits_{c-i\infty}^{c+i\infty}\frac{\text{d}s}{2\pi i}  b_T^{-s} \widetilde{\Gamma}_j(\mu_0,s)\mathcal{E}_{ji}(\mu,\mu_0;s+3)\ , %
\end{equation}
where we replace $N$ with $s$, assuming an analytical continuation on the complex plane has been performed.

\begin{figure}[t]
  \begin{center}
   \includegraphics[width=0.483\textwidth]{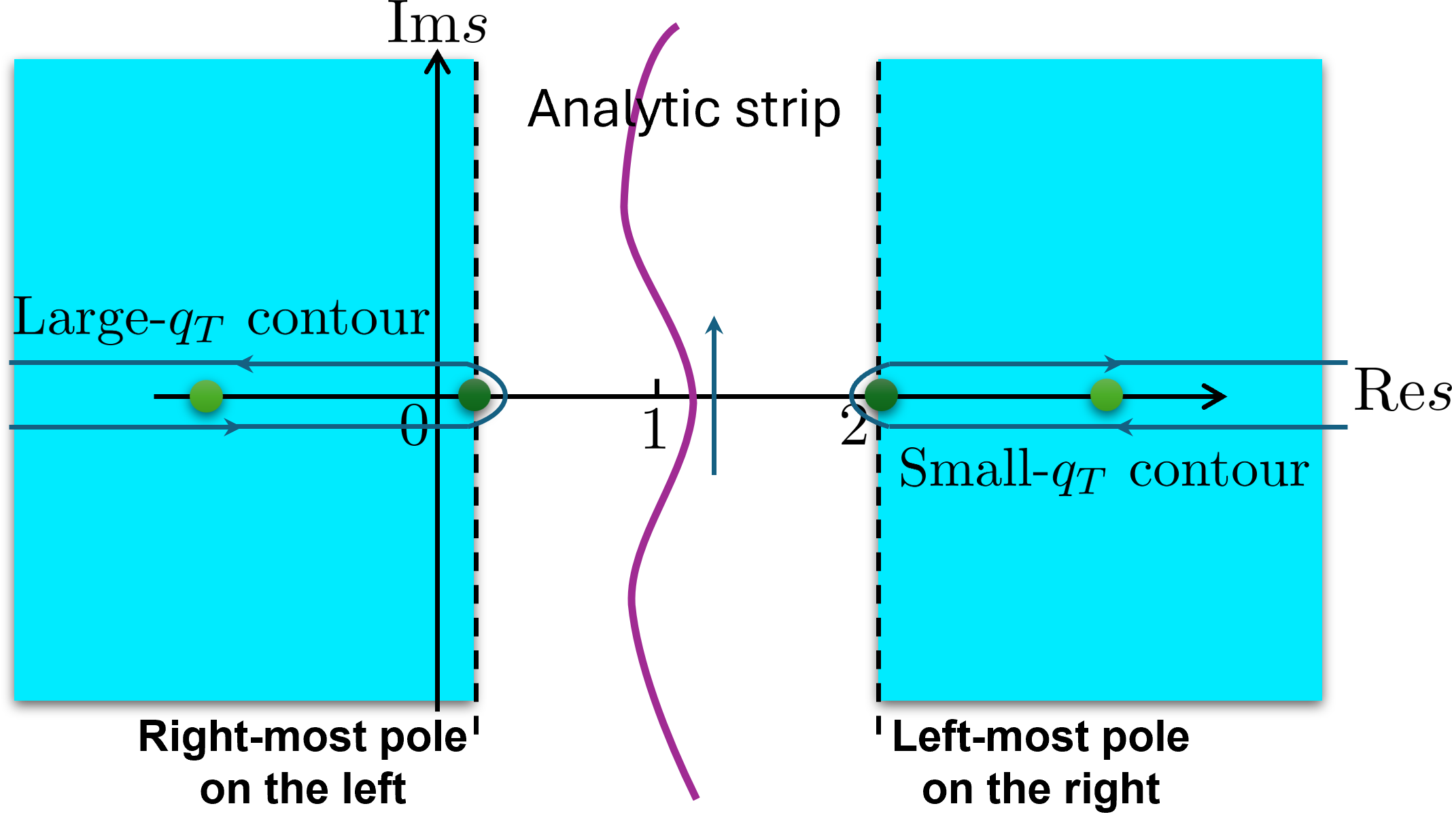} 
\caption{Singularity structures and contours on the complex $s$ plane for the inverse Mellin transformation. These singularity structures have also been discussed in the context of light-ray OPE~\cite{Chen:2024nyc, Chang:2025kgq}.}
  \label{fig:Mellincontour}
 \end{center}
  \vspace{-5.ex}
\end{figure}

One prominent property of the Mellin-space representation of the unintegrated EEC jet functions, beyond its multiplicative form and straightforward exponentiation solution, is its ability to naturally encode the distinct scaling and evolution behaviors at both small and large $q_T$~\cite{Guo:2025qnz}, since it is an exact solution of the evolution equation in both regions.
To see this more clearly, we can Fourier transform the above equation into $q_T$-space,
\begin{eqnarray}
    \label{eq:mellinevoqTfinal}
   {\Gamma}_i(\mu,q_T)&=&\frac{1}{q_T^2} %
\int\limits_{c-i\infty}^{c+i\infty}\frac{\text{d}s}{2\pi i}\left(\frac{q_T}{2}\right)^{s}\frac{\Gamma(1-s/2)}{\pi\Gamma(s/2)}\nonumber\\
    &&\times 
    \widetilde{\Gamma}_j(\mu_0,s){\cal E}_{ji}(\mu,\mu_0;s+3) \ .  
 \end{eqnarray}
Assuming that the only singularities in the complex plane are poles on the real axis as shown in Fig.~\ref{fig:Mellincontour}, the residue theorem implies schematically the following,
\begin{equation}
\label{eq:residualtheorem}
    \Gamma_i(\mu,q_T)\sim \frac{1}{q_T^2}\sum_{k}  r_k \, q_T^{N_k}\, {\cal E}_{ji}\left(\mu,\mu_0;N_k+3\right)\, ,
\end{equation}
where $N_k$ corresponds to the location of the $k$-th pole of the integrand, and $r_k$ denotes the associated residue. Here, we make the scale-dependent evolution operator explicit for clarity and ignore its singularities.

{This representation makes explicit that the scaling behavior of the unintegrated EEC jet functions is governed by the pole structures, establishing a natural correspondence with the scaling behavior of DGLAP operators in the light-ray OPE framework~\cite{Chen:2024nyc, Chang:2025kgq}.}
Particularly, in the small-$q_T$ limit, the pole at $s=2$ yields a contribution proportional to $q_T^2$, which cancels the overall prefactor, leading to a constant behavior for the EEC jet function, $\Gamma_i(q_T)\big|_{q_T\to 0}\sim \mathrm{const.}$, with $\mu$-dependence governed by the anomalous dimension $\gamma^{(5)}_{ji}$~\cite{Chang:2025kgq}. On the other hand, in the large-$q_T$ regime, the pole near $s=0$ arises from the initial condition of $\widetilde{\Gamma}_j(\mu_0,s)$ and gives rise to $\Gamma_i(q_T)\big|_{q_T\gg \Lambda_i}\sim 1/q_T^2$, with the corresponding $\mu$-dependence controlled by $\gamma^{(3)}_{ji}$, where $\Lambda_i$ represents the transition scale between these two regions. 
An appropriate determination of the initial condition $\widetilde{\Gamma}_j$ is required to be consistent with the collinear scaling of the unintegrated EECs at large $q_T$~\cite{Dixon:2019uzg}, as discussed below.

\section{Determining the Initial Condition and Comparing to Collinear Scalings at Large $q_T$}

The initial condition of $\widetilde{\Gamma}_i(\mu_0,b_T)\longrightarrow \widetilde{\Gamma}_i(\mu_0,s)$ is an important ingredient to apply the Mellin-space approach, where several considerations have to be taken into account to construct it. First, we need to include a non-perturbative form factor $\Lambda_i$ as an IR regulator to suppress the large $b_T$ contributions. Second, we need to incorporate the collinear scaling behavior at small $b_T$~\cite{Guo:2025zwb, Guo:2025qnz} that is consistent with the scaling behavior at large $q_T$~\cite{Dixon:2019uzg}. 
To do that, we start with the following ansatz in the fixed-coupling limit,
\begin{equation}
    \widetilde{\Gamma}^{\rm{fixed}}_{i}(\mu,b_T)=\widetilde{\Gamma}^0_j \exp\left[-\gamma^{\mathrm{FP}}_{ji} \ln(\mu^2/\mu_b^2)\right]\ ,
\end{equation}
where the exponential should be understood as matrix exponents. It is straightforward to verify that the evolution Eq.~(\ref{eq:evolutionbt}) implies the following for the fixed-point (FP) anomalous dimension $\gamma_{ji}^{\rm{FP}}$ that,
\begin{eqnarray}
    \gamma_{ji}^{\rm{FP}}&=&- \int{\dd x}\  x^{2+2\gamma_{jk}^{\rm{FP}}} \hat{P}_{ki}(x) \ .
\end{eqnarray}
This ansatz reproduces the collinear results order by order up to $\beta$-function terms associated with running coupling. This behavior of EEC jet functions with fixed coupling has been discussed in more detail in~\cite{Dixon:2019uzg, Lee:2024icn} in the context of the reciprocity relation in conformal field theory. The fixed-point solution of the anomalous dimensions is nothing but the space-like DGLAP anomalous dimensions at $N=3$~\cite{Mueller:1983js, Dokshitzer:2005bf, Marchesini:2006ax, Basso:2006nk, Dokshitzer:2006nm}. This relation appears to still hold in QCD, examined up to NNLO~\cite{Chen:2020uvt}. We also note that, throughout this work, we perturbatively expand the anomalous dimensions as $\gamma_{ji}^{(s)}=\sum a_s^n \gamma_{n, ji}^{(s)}$ with $a_s\equiv\alpha_S/(2\pi)$, which have been calculated up to NNLO in the literature~\cite{Mitov:2006ic, Moch:2007tx}, as well as the QCD $\beta$ function $\beta(\mu)\equiv-\text{d}\ln a_S(\mu)/\text{d}\ln\mu^2$ as $\beta(\mu)=\sum a_s^{n+1}(\mu) \beta_n$. 

The above fixed-coupling ansatz allows one to write down the following ansatz for running coupling:
\begin{eqnarray}
\label{eq:fpansatz}
    \widetilde{\Gamma}_{i}(\mu,b_T)&=&\widetilde{\Gamma}^0_j(\mu_b) \text{P}\exp\left[-\int_{\mu^2_b}^{\mu^2}\frac{\text{d}\mu'^2}{\mu'^2}\gamma^{\mathrm{FP}}_{ji}(a_s(\mu'))  \right]\ ,\nonumber\\
    &=&\widetilde{\Gamma}^0_j(\mu_b) {\cal E}^S_{ji}(\mu,\mu_b;N=3)\ ,
\end{eqnarray}
where we used the reciprocity relation in the second line to rewrite it as the space-like DGLAP evolution kernel ${\cal E}^S_{ji}$ at $N=3$. Here we neglect the running-coupling modification to $\gamma^{\rm{FP}}$ , i.e., the $\beta$-function term in Eq. (45) of~\cite{Dixon:2019uzg}. To regulate the Landau pole at large $b_T$, We use the $b_*$ prescription that $\mu_{b}\equiv2e^{-\gamma_E}/b_*$ with $b_*\equiv b_T/\sqrt{1+b_T^2/b_{\textrm{max}}^2}$ and set $b_{\textrm{max}}=1.5~\rm GeV^{-1}$. This ansatz ensures the boundary condition $\widetilde{\Gamma}_{i}(\mu=\mu_b,b_T)=\widetilde{\Gamma}_i^{0}(\mu_b)$ as required in TMD-type resummation, which consists of both perturbative and non-perturbative parts,
\begin{equation}
\label{eq:nonpertboundary}
    \widetilde{\Gamma}_i^{0}(\mu_b)
=N_ie^{-\Lambda_i b_T}\left[j_i^{(0)}+a_s(\mu_b)j_i^{(1)}+a_s^2(\mu_b)j_i^{(2)}\right]\ ,
\end{equation}
up to NNLO~\cite{Dixon:2019uzg}. The normalization $N_{q/g}$ is related to the integrated EEC jet functions to be specified later. 

\begin{figure}[tb]
  \begin{center}
   \includegraphics[width=0.483\textwidth]{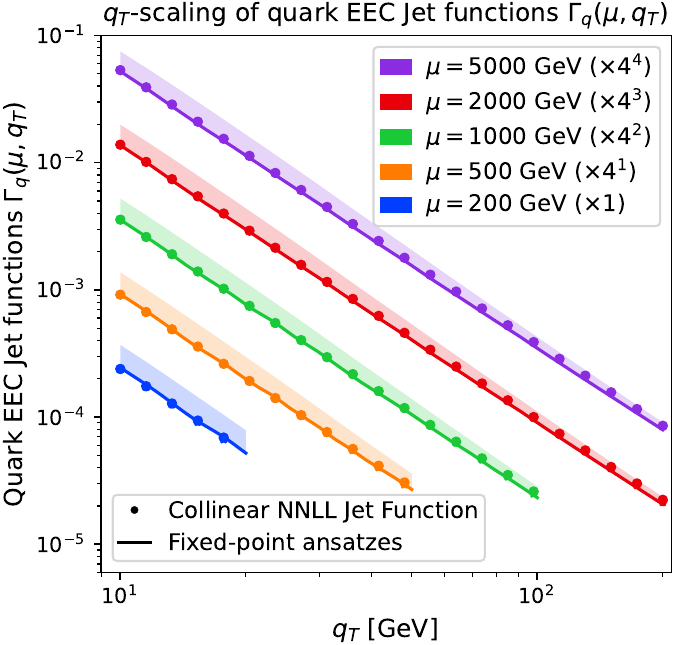} 
\caption{Comparison with the derivative of the cumulant EEC jet functions from collinear calculations. The fixed-point ansatz, using $\Lambda_q=\Lambda_g=0.5$ GeV and $\mu_0=10$ GeV, shows excellent agreement in the perturbative regime, with similar agreement for the gluon. Bands from varying $\Lambda_q=\Lambda_g$ between 0.5 and 3 GeV indicate sizable power corrections at lower $q_T$.}
  \label{fig:qTscalingcomp}
 \end{center}
  \vspace{-5.ex}
\end{figure}

In this work, we compute $\widetilde{\Gamma}_{i}(\mu,b_T)$ and its Mellin moment $\widetilde{\Gamma}_{i}(\mu,s)$ at the initial scale $\mu_0$ numerically. We also explicitly verify that the Mellin moment $\widetilde{\Gamma}_{i}(\mu_0,s)$ reproduces the corresponding $b_T$-space ansatz through inverse Mellin transformation with sub-percent accuracy. We then perform the full NNLO evolution with running coupling from $\mu_0$ to an arbitrary scale $\mu$ using the Mellin-space method described above. The unintegrated EEC jet functions at scale $\mu$ can be reconstructed afterward in either $b_T$ or $q_T$ space. We emphasize that the approximate ansatz is introduced solely to define the initial condition at $\mu_0$ with the appropriate perturbative input. This Mellin-space framework, on the other hand, is applicable to general-purpose evolutions of unintegrated EEC jet functions in future studies.

In Fig.~\ref{fig:qTscalingcomp}, we compare the quark unintegrated EEC jet functions obtained from this ansatz with those derived from collinear calculations. The latter are obtained by differentiating the cumulant jet functions of~\cite{Dixon:2019uzg} with respect to $z$. For this comparison, we choose an initial scale $\mu_0=10$~GeV and non-perturbative parameters $\Lambda_q=\Lambda_g=0.5$ GeV, and set $N_q=N_g=1$ to match perturbative normalization. We find excellent agreement between the fixed-point ansatz and the collinear results in the regime $q_T\gg\Lambda$, while sizable power corrections arise at lower $q_T$, particularly for larger values of $\Lambda_{q,g}$. We also observe that the unintegrated EEC jet functions exhibit a modified scaling behavior, $q_T^{-2}\to q_T^{-2-\delta}$.

\section{EECs in $e^+e^-$ Annihilation at NNLO}
The developments above can be easily applied to the EECs as well. By introducing the Mellin moments of the $b_T$-space EEC $\widetilde{\Sigma}^{e^+e^-}_2(Q,b_T)$ with respect to $b_T$ and those of the hard coefficients with respect to $x$ as,
\begin{eqnarray}
    \widetilde{\Sigma}^{e^+e^-}_2(Q,N)&\equiv&\int \text{d}b_T\, b_T^{N-1} \widetilde{\Sigma}^{e^+e^-}_2(Q,b_T)\ ,\\
    H_i^{e^+e^-}(Q/\mu,N)&\equiv&\int \text{d}x\, x^{N-1}H^{e^+e^-}_i(x,Q/\mu)\ ,
\end{eqnarray}
the factorization formula of $\widetilde{\Sigma}^{e^+e^-}_2(Q,b_T)$ in Eq. (\ref{eq:factorizationbtee}) has a multiplicative form in Mellin space:
\begin{equation}
    \widetilde{\Sigma}^{e^+e^-}_2(Q,N)= H^{e^+e^-}_i(Q/\mu,N+3)\,\widetilde{\Gamma}_i(\mu,N)\ ,
\end{equation}
in terms of the Mellin moments of unintegrated EEC jet functions $\widetilde{\Gamma}(\mu, N)$ defined above. 

Then we arrive at the following final result for the EECs at small angles in $e^+e^-$ annihilation,
\begin{align}
\begin{split}
     &\left.2\frac{\text{d}\Sigma_2^{e^+e^-}}{\pi Q^2\text{d} z}\right|_{z\ll 1}=\frac{\sigma_0}{\sigma_{tot}}\frac{1}{q_T^2} \int\limits_{c-i\infty}^{c+i\infty}\frac{\text{d}s}{2\pi i} \left(\frac{q_T}{2}\right)^{s}\frac{\Gamma(1-s/2)}{\pi\Gamma(s/2)}\\
    &\quad\quad \times \widetilde{\Gamma}_j^{\mathrm{Init}}(\mu_0,s)
    {\cal E}_{ji}(\mu_F,\mu_0;s+3) H^{e^+e^-}_i\left({Q}/{\mu_F},s+3\right)\ .  
\end{split}
\end{align}
We also include the non-perturbative normalization $N_i=\Gamma_i(\mu_b)$ in the initial condition and refer to the corresponding moment-space ansatz as $\widetilde{\Gamma}_j^{\mathrm{Init}}(\mu_0,s)$. The integrated EEC jet functions $\Gamma_{j}(\mu)$ are related to the $z^2$ moment of the single-hadron fragmentation functions $\Gamma_i'(\mu)$ via $\Gamma_i(\mu)=1-\Gamma_i'(\mu)$~\cite{Guo:2025qnz}. %
Specifically, we use the fragmentation functions from~\cite{Gao:2025hlm}, which give $\Gamma'_q(\mu'_0)=0.256$ and $\Gamma'_g(\mu_0')=0.130$ at $\mu'_0=2$ GeV. The $\Gamma_i(\mu_b)$ is then obtained by evolving them to the scale $\mu_b$ with the corresponding time-like anomalous dimensions.

\begin{figure}[t]
  \begin{center}
   \includegraphics[width=0.483\textwidth]{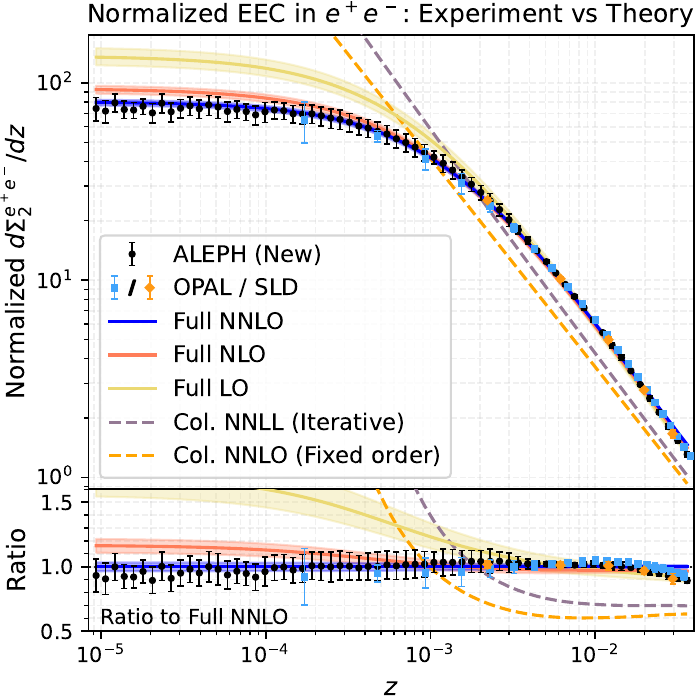} 
\caption{Comparison to new analysis of ALEPH data on charged-hadron EECs~\cite{Electron-PositronAlliance:2025fhk} and the earlier OPAL~\cite{OPAL:1991uui} and SLD~\cite{SLD:1994idb} data at $Q=91.2$ GeV, with $\Lambda_q=\Lambda_g=2.5~\text{GeV}$ and $\Gamma'(\mu)$ from fragmentation functions~\cite{Gao:2025hlm}. Collinear calculations taken from~\cite{Dixon:2019uzg}. Experiment data are scaled by $1/N_{\rm{data}}$ so that their integrals reproduce the integrated EEC as discussed in the text~\cite{Guo:2025qnz}. Bands obtained from varying the factorization scale $\mu_F$ by a factor of $1/2$ to $2$.} \label{fig:comparison}
 \end{center}
  \vspace{-5.ex}
\end{figure}

\begin{figure}[t]
  \begin{center}
   \includegraphics[width=0.483\textwidth]{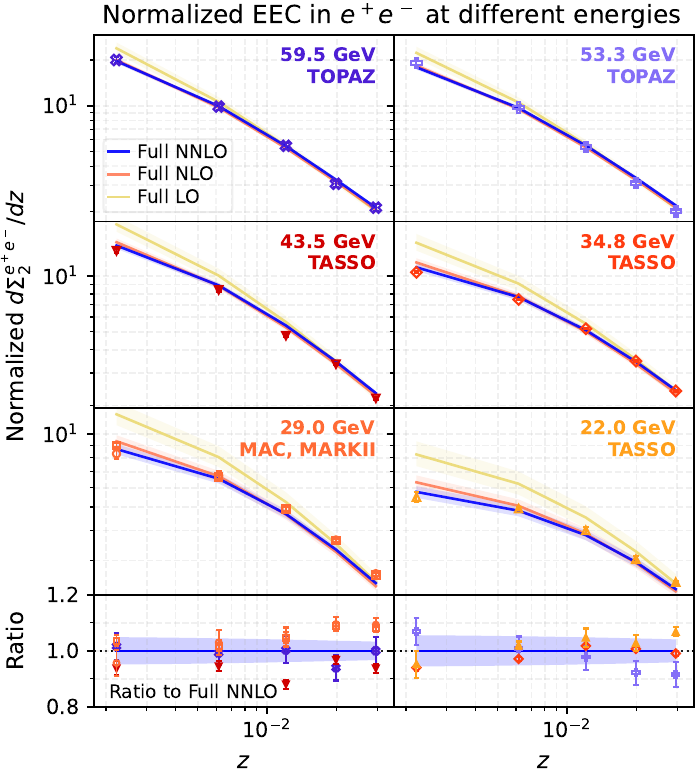} 
\caption{Comparison to EEC measurements across lower energies: $Q=29.0$ GeV by MAC~\cite{Fernandez:1984db} and MARK II~\cite{Wood:1987uf}, $Q=22.0$, 34.8, and 43.5 GeV by TASSO~\cite{TASSO:1987mcs} and $Q=53.3$ and 59.5 GeV by TOPAZ~\cite{TOPAZ:1989yod}. Setup and parameters are identical to those of Fig.\ref{fig:comparison}. \label{fig:comparison2}}
 \end{center}
  \vspace{-5.ex}
\end{figure}

We compare our theoretical predictions with the recent analysis of ALEPH data on charged-hadron EECs~\cite{Electron-PositronAlliance:2025fhk} together with earlier OPAL~\cite{OPAL:1991uui} and SLD~\cite{SLD:1994idb} measurements at $Q=91.2~\mathrm{GeV}$ in Fig.~\ref{fig:comparison} and at lower energies by MAC~\cite{Fernandez:1984db}, MARK II~\cite{Wood:1987uf}, TASSO~\cite{TASSO:1987mcs} and TOPAZ~\cite{TOPAZ:1989yod} in Fig.~\ref{fig:comparison2}. %
{The quoted Full LO/NLO/NNLO theory accuracy refers to a consistent treatment of perturbative expansions at the corresponding order, of the hard coefficients $H_i$, jet function constants $j_i$, anomalous dimensions $\gamma^T$ entering the evolution operators, and $\gamma^S$ in the initial ansatz, as well as the running coupling.} As discussed in~\cite{Guo:2025qnz}, the integrated EEC $\Sigma_2^{e^+e^-}(Q)$ is given by the integrated EEC jet function $\Gamma_i(Q)$. We therefore determine the data normalization via $\int \mathrm{d}\Sigma_2^{e^+e^-} = N_{\mathrm{Data}}\,\Sigma_2^{e^+e^-}(Q)$, using the $\Sigma_2^{e^+e^-}(Q)$ from~\cite{Guo:2025qnz}. For ALEPH, we find $N_{\mathrm{ALEPH}} \approx (2/3)^2/2 \times 0.87$, where the first two factors account for the charged-particle measurement and the removal of pair double counting, while the final factor is determined numerically and likely reflects non-hadronic final-state contributions. For the remaining datasets, we have, ignoring the weak $Q$ dependence, $N=1/\Sigma_2^{e^+e^-}(Q)\approx 1.1$, since these datasets were originally normalized to unity upon integration. The bins with the smallest angles in these datasets appear anomalous and are therefore removed for comparison.

In this setup, the only additional non-perturbative inputs are the scales $\Lambda_q$ and $\Lambda_g$, which we take to be $\Lambda_q=\Lambda_g=2.5~\text{GeV}$ to reproduce the small-$q_T$ behavior of the data. We emphasize that the $e^+e^-$ annihilation data are not sensitive to the gluon EEC jet function. In particular, we find that a wide range of $\Lambda_g$ leads to almost the same results for Fig.~\ref{fig:comparison} and ~\ref{fig:comparison2}. To constrain $\Lambda_g$, or in general, the non-perturbative gluon EEC jet functions, we need to compare to the EEC measurements at hadron colliders. For the theoretical uncertainties in this paper, we set the central value of the factorization scale to $\mu_F=Q$ and %
vary it by a factor of $1/2$ to $2$.

In general, the full NNLO theory predictions agree very well with the data across the full kinematic and energy range of nearside with small angles $\chi\le0.4$ and $22.0\le Q\le91.2$ GeV. Second, the theoretical uncertainties become smaller with higher-order corrections included. Third, even in the relatively moderate-$z$ region where we expect perturbative contributions to dominate, full resummation with non-perturbative input improves the agreement between the theory and experiment. This indicates the importance of power corrections, as also shown in Ref.~\cite{Chen:2024nyc,Lee:2024esz}, whereas the resummation formula takes into account these contributions in a consistent fashion. %
Finally, we note that in the small-$z$ region the sizable contributions from higher order terms point to a potential need of additional resummation which will help improve the perturbative convergence as well. We will come back to this issue in a future publication.

\section{Conclusion}

In summary, we have established the Mellin-transform method as a precision framework for nearside EEC analyses. We solve the evolution equation analytically in the Mellin space and take into account the NNLO perturbative QCD corrections for the nearside EECs in $e^+e^-$ annihilation with NNLL resummation of all-order collinear gluon contributions. The final results depend only on one parameter, $\Lambda$, which characterizes the transition scale between the perturbative and post-confinement regions of the nearside EECs, and show excellent agreements with existing measurements of nearside EEC.

{The results in Figs.~\ref{fig:comparison} and~\ref{fig:comparison2} also underscore the importance of power corrections even in the collinear perturbative region, which are efficiently incorporated within this framework.}
This will become an important ingredient to determine the running coupling constant from future measurements in $e^+e^-$ annihilation. 

{The presented benchmark establishes a compelling basis for future applications along the direction outlined in this work.}
As mentioned above, the gluon EEC jet function is not well constrained from $e^+e^-$ data. The EECs in jets at the LHC will provide important constraints on the gluon EECs. Our method can be extended to those measurements. In addition to the two-point energy correlators, extension to three-point energy correlators shall be carried out as well. This will help to improve the precision for the $\alpha_s$ determination from the EEC measurements at the LHC. We plan to carry out these studies in a future publication. 

\section*{Acknowledgment}

We thank Hao Chen and Werner Vogelsang for their collaborations at the early stage of this work and for their valuable discussions and comments. %
We also thank Felix Ringer and Wenbin Zhao for useful correspondence. We used the \texttt{EKO} package for the NNLO anomalous dimension and running coupling constant~\cite{Candido:2022tld}. The hard coefficients of single inclusive hadron production in $e^+e^-$ annihilation are taken from~\cite{Mitov:2006wy}. We also use the \texttt{HarmonicSums} package~\cite{Ablinger:2014rba} to cross-check some moment-space expressions and thank Carsten Schneider for help with accessing the package. This work is supported by the Office of Science of the U.S. Department of Energy under Contract No. DE-AC02-05CH11231. Y. G. is also supported by the U.S. Department
of Energy under Contract No. DE-SC0012704, and by LDRD funds from Brookhaven Science Associate.

\bibliographystyle{apsrev4-1}
\bibliography{refs.bib}

\end{document}